# Cosmic censorship and stationary states of half-spin particles in the field of Reissner-Nordström naked singularity

M.V.Gorbatenko[1], V.P.Neznamov[1,2*], E.Yu.Popov[1], I.I.Safronov[1]

[1]RFNC-VNIIEF, Russia, Sarov, Mira pr., 37, 607188
[2]National Research Nuclear University MEPhI, Moscow, Russia

Abstract

The paper explores quantum mechanics of half-spin particle motion in the field of Reissner-Nordström (RN) naked singularity. It is shown that for any quantum mechanical Dirac particle, irrespective of availability and sign of its electrical charge, the RN naked singularity is separated by an infinitely high positive potential barrier. With like charges of a particle and the source of the RN naked singularity, near the origin there exists the second completely impenetrable potential barrier.

It has been proved that in the field of the RN naked singularity, bound states of half-spin particles can exist. The conditions for appearance of such states were revealed and computations were performed to find energy eigenvalues and eigenfunctions.

---

[*] E-mail: neznamov@vniief.ru

# Introduction

Among singular solutions to the general relativity theory, there exist black holes with event horizons and naked singularities. The hypothesis of cosmic censorship proposed more than 40 years ago [1] bans the existence of singularities not shielded by event horizons. However, there is still no complete proof of this hypothesis. Along with black holes, many researchers consider the formation of naked singularities, their stability and distinctive features during experimental observations [2] -[7].

Horowitz and Marolf have shown [8] that there exist static metrics with timelike curvature singularities which appear completely nonsingular when probed with quantum test spinless particles. In our work we have confirmed the results [8] as applied to the motion of quantum-mechanical half-spin particles in the field of Reissner-Nordström (RN) naked singularity. In the paper, for this case, we also consider the possibility of existence of stationary bound state half-spin particles. The analysis was performed both by means of effective potentials in the one-dimensional Schrödinger-type equation and by direct solution to the Dirac equation in the field of RN naked singularity. The method of effective potentials as applied to the motion of a half-spin particle in Schwarzschild and RN fields is explored in more detail in [9]. In this paper physical characteristics of half-spin particle interaction with the field of RN naked singularity are also discussed.

Earlier in [10], C.L.Pekeris, K.Frankowski studied the solution to the Dirac equation in the field of RN naked singularity with the phenomenological introduction of Kerr-Newman geometry into the neighborhood of the origin. Their effort was aimed at determining the extremely low gravitational shift of energy levels of hydrogen atoms. In our work, we used their boundary conditions for wave functions at the origin of the RN geometry.

The qualitative conclusion of existence of stationary bound states of Dirac particles in the field of RN naked singularity was made earlier by V. Dzhunushaliev in [11]. However, the absence of unambiguity in the boundary condition for wave functions at the origin and the equations used for radial functions proposed in [11] did not allow using the computational method proposed by the author for determination of the discrete energy spectrum of Dirac particles.

In section 2, the Reissner-Nordström metric, the self-conjugate Dirac Hamiltonian in the RN field, the system of equations for radial wave functions are given for reminding basic notions and facts.

In section 3, the obtained effective potentials of the one-dimensional Schrödinger-type equation for the field of RN naked singularity are analyzed.

In section 4, physical characteristics of half-spin particle interaction with the field of RN naked singularity are discussed.

In section 5, stationary bound states of Dirac particles in the field of RN naked singularity are determined.

In section 6, the obtained results are discussed.

In the paper, as a rule, we use system of units $\hbar = c = 1$; the space signature is chosen to be

$$g_{\alpha\beta} = diag[1,-1,-1,-1]. \tag{1}$$

## 2. The Reissner-Nordström metric

The line element of the Reissner-Nordström metric is

$$ds^2 = f_{R-N}dt^2 - \frac{dr^2}{f_{R-N}} - r^2\left(d\theta^2 + \sin^2\theta d\varphi^2\right), \tag{2}$$

where $f_{R-N} = \left(1 - \frac{r_0}{r} + \frac{r_Q^2}{r^2}\right)$, $r_0 = \frac{2GM}{c^2}$ is the gravitational radius of the Schwarzschild field, $r_Q = \frac{\sqrt{G}Q}{c^2}$ is the "charge" radius, $G$ is the gravitational constant, $c$ is the velocity of light.

1. If $r_0 > 2r_Q$, then

$$f_{R-N} = \left(1 - \frac{r_+}{r}\right)\left(1 - \frac{r_-}{r}\right), \tag{3}$$

where $r_\pm$ are the radii of the outer and inner event horizons

$$r_\pm = \frac{r_0}{2} \pm \sqrt{\frac{r_0^2}{4} - r_Q^2}. \tag{4}$$

2. The case $r_0 = 2r_Q$ corresponds to the Reissner-Nordström extreme field.

3. The case $r_0 < 2r_Q$ corresponds to the naked singularity. In this case, there is always $f_{R-N} > 0$ and the domain of wave functions is the entire region $r \in [0, \infty)$.

Below, we will analyze the behavior of effective potentials of the Dirac equation in the field of RN naked singularity.
For obtaining one-dimensional the Schrödinger-type equation with the self-conjugate Hamiltonian the initial Dirac Hamiltonian must be also self-conjugate.

The algorithms for deriving self-conjugate Dirac Hamiltonians in external gravitational fields by means of the methods of pseudo-Hermitian quantum mechanics are represented in [12] - [14].

The self-conjugate Hamiltonian of a massive and charged half-spin particle in the RN field was obtained in [15]

$$H_\eta = \sqrt{f_{R-N}}\beta m - i\alpha^1\left(f_{R-N}\frac{\partial}{\partial r} + \frac{1}{r} - \frac{r_0}{2r^2}\right) - \\ -i\sqrt{f_{R-N}}\frac{1}{r}\left[\alpha^2\left(\frac{\partial}{\partial \theta} + \frac{1}{2}\text{ctg}\theta\right) + \alpha^3\frac{1}{\sin\theta}\frac{\partial}{\partial \varphi}\right] + \frac{eQ}{r}. \tag{5}$$

In (5), $\alpha^k, \beta$ are Dirac matrices.

After separation of variables we obtain the system of equations for radial functions $F_{R-N}(\rho), G_{R-N}(\rho)$

$$f_{R-N}\frac{dF_{R-N}(\rho)}{d\rho} + \left(\frac{1+\kappa\sqrt{f_{R-N}}}{\rho} - \frac{\alpha}{\rho^2}\right)F_{R-N}(\rho) - \left(\varepsilon - \frac{\alpha_{em}}{\rho} + \sqrt{f_{R-N}}\right)G_{R-N}(\rho) = 0, \\ f_{R-N}\frac{dG_{R-N}(\rho)}{d\rho} + \left(\frac{1-\kappa\sqrt{f_{R-N}}}{\rho} - \frac{\alpha}{\rho^2}\right)G_{R-N}(\rho) + \left(\varepsilon - \frac{\alpha_{em}}{\rho} - \sqrt{f_{R-N}}\right)F_{R-N}(\rho) = 0, \tag{6}$$

In (6) we use dimensionless variables

$$\rho = \frac{r}{l_c};\ \varepsilon = \frac{E}{m};\ \alpha = \frac{r_0}{2l_c} = \frac{GMm}{\hbar c} = \frac{Mm}{M_P^2}, \\ \alpha_Q = \frac{r_Q}{l_c} = \frac{\sqrt{G}Qm}{\hbar c} = \frac{\sqrt{\alpha_{fs}}}{M_P}m\frac{Q}{|e|};\ \alpha_{em} = \frac{eQ}{\hbar c} = \alpha_{fs}\frac{Q}{e} \tag{7}$$

$l_c = \frac{\hbar}{mc}$ is the Compton wave-length of a Dirac particle; $E$ is its energy;

$M_P = \sqrt{\frac{\hbar c}{G}} = 2.2\cdot 10^{-5}g\ (1.2\cdot 10^{19} GeV)$ is the Planck mass; $\alpha_{fs} \approx \frac{1}{137}$ is the electromagnetic fine structure constant; $\alpha, \alpha_{em}$ are gravitational and electromagnetic coupling constants; $\alpha_Q$ is the dimensionless constant characterizing source of the electromagnetic field in the Reissner-Nordström metric;

The separation constant $\kappa = \mp\left(j+\frac{1}{2}\right) = \begin{cases} -(l+1), & j = l+\frac{1}{2} \\ l, & j = l-\frac{1}{2} \end{cases}$ ; $l, j$ are quantum numbers of orbital and total angular momenta of a Dirac particle.

## 3. Effective potentials for the field of Reissner-Nordström naked singularity

From the system of equations (6), we will derive the second-order equation for the function $\psi(\rho)$ proportional either to $F(\rho)$ or to $G(\rho)$

In the first case,

$$\psi(\rho) = F(\rho) \exp\left(\frac{1}{2}\int A_1(\rho')d\rho'\right). \tag{8}$$

The equation for $\psi(\rho)$ has the form of the nonrelativistic Schrödinger equation

$$\frac{d^2\psi(\rho)}{d\rho^2} + 2\left(E_{schr} - U_{eff}(\rho)\right)\psi(\rho) = 0. \tag{9}$$

In equation (9),

$$E_{schr} = \frac{1}{2}(\varepsilon^2 - 1),$$

$$U_{eff}(\rho) = E_{schr} + \frac{1}{4}\frac{dA_1(\rho)}{d\rho} + \frac{1}{8}A_1^2(\rho) - \frac{1}{2}B_1(\rho). \tag{10}$$

In expressions (8), (10),

$$A_1(\rho) = -\frac{1}{B(\rho)}\frac{dB(\rho)}{d\rho} - A(\rho) - D(\rho),$$

$$B_1(\rho) = -B(\rho)\frac{d}{d\rho}\left(\frac{A(\rho)}{B(\rho)}\right) - C(\rho)B(\rho) + A(\rho)D(\rho). \tag{11}$$

In expressions (11),

$$A(\rho) = -\frac{1}{f_{R-N}}\left(\frac{1+\kappa\sqrt{f_{R-N}}}{\rho} - \frac{\alpha}{\rho^2}\right),$$

$$B(\rho) = \frac{1}{f_{R-N}}\left(\varepsilon - \frac{\alpha_{em}}{\rho} + \sqrt{f_{R-N}}\right),$$

$$C(\rho) = -\frac{1}{f_{R-N}}\left(\varepsilon - \frac{\alpha_{em}}{\rho} - \sqrt{f_{R-N}}\right),$$

$$D(\rho) = -\frac{1}{f_{R-N}}\left(\frac{1-\kappa\sqrt{f_{R-N}}}{\rho} - \frac{\alpha}{\rho^2}\right). \tag{12}$$

The effective potentials $U_{eff}(\rho,\kappa,\alpha,\alpha_Q,\alpha_{em},\varepsilon)$, determined by the expressions (10) - (12), have rather a cumbersome form and have been calculated in this paper due to the "Maple". Note that the energy of a Dirac particle in the expressions (10) - (12) is one of the parameters. The value $\varepsilon < 1$ corresponds to bound states of a half-spin particle. Let us consider some peculiar features of the expressions $U_{eff}(\rho,\kappa,\alpha,\alpha_Q,\alpha_{em},\varepsilon)$.

**3.1** Irrespective of signs of charges $Q, e$, the leading term of the effective potential near the origin is

$$U_{eff} = \frac{3}{8\rho^2} + O\left(\frac{1}{\rho}\right) \text{ at } \rho \to 0, \qquad (13)$$

i.e., the Reissner-Nordström singularity is separated from Dirac particles by an infinitely large effective positive barrier. This agrees with the results [8] as applied to the motion of spinless particles in the fields of certain singular metrics.

**3.2** With like signs of charges $Q, e$ near the origin there is the second potential barrier of the form

$$U_{eff}\bigg|_{\rho \to \rho_{cl}} = \frac{1}{4} \frac{1}{\left(\varepsilon - \frac{\alpha_{em}}{\rho} + \sqrt{1 - \frac{2\alpha}{\rho} + \frac{\alpha_Q^2}{\rho^2}}\right)^2} \left(\frac{\alpha_{em}}{\rho^2} + \frac{\frac{\alpha}{\rho^2} - \frac{\alpha_Q^2}{\rho^3}}{\sqrt{1 - \frac{2\alpha}{\rho} + \frac{\alpha_Q^2}{\rho^2}}}\right)^2 + O\left(\frac{1}{\varepsilon - \frac{\alpha_{em}}{\rho} + \sqrt{1 - \frac{2\alpha}{\rho} + \frac{\alpha_Q^2}{\rho^2}}}\right). \qquad (14)$$

The radius $\rho_{cl}$, at which the expression in the denominator of the first summand in (14) is equal to zero, is determined by the following equation

$$\rho_{cl} = \frac{\alpha_{em}\varepsilon - \alpha - \sqrt{(\alpha_{em}\varepsilon - \alpha)^2 - (\alpha_{em}^2 - \alpha_Q^2)(\varepsilon^2 - 1)}}{\varepsilon^2 - 1}. \qquad (15)$$

For all the cases considered by the authors (see below item 4), the coefficient at the leading singularity in (14)

$$A = \frac{1}{4}\left(\frac{\alpha_{em}}{\rho_{cl}^2} + \frac{\frac{\alpha}{\rho_{cl}^2} - \frac{\alpha_Q^2}{\rho_{cl}^3}}{\sqrt{1 - \frac{2\alpha}{\rho_{cl}} + \frac{\alpha_Q^2}{\rho_{cl}^2}}}\right)^2 > \frac{3}{8}. \qquad (16)$$

It is well-known that such potential barriers are impenetrable to quantum-mechanical particles [16].

If $\alpha_{em} \gg \alpha$ and $\alpha_{em} \gg |\alpha_Q|$, then the expression (15) turns into

$$\rho_{cl} = \frac{\alpha_{em}}{\varepsilon + 1}. \qquad (17)$$

In this case, the value $r_{cl} = l_c \rho_{cl}$ is proportional to the value of the classical radius of a particle with mass $m$ and charge $e$ interacting with the potential $\frac{Q}{r}$

$$r_{cl} = \frac{eQ}{mc^2} \frac{1}{1 + \frac{E}{mc^2}}. \tag{18}$$

Fig.1 presents the typical form of $U_{eff}(\rho)$ for the case of like signs of charges $Q, e$ and $\varepsilon > 1$.

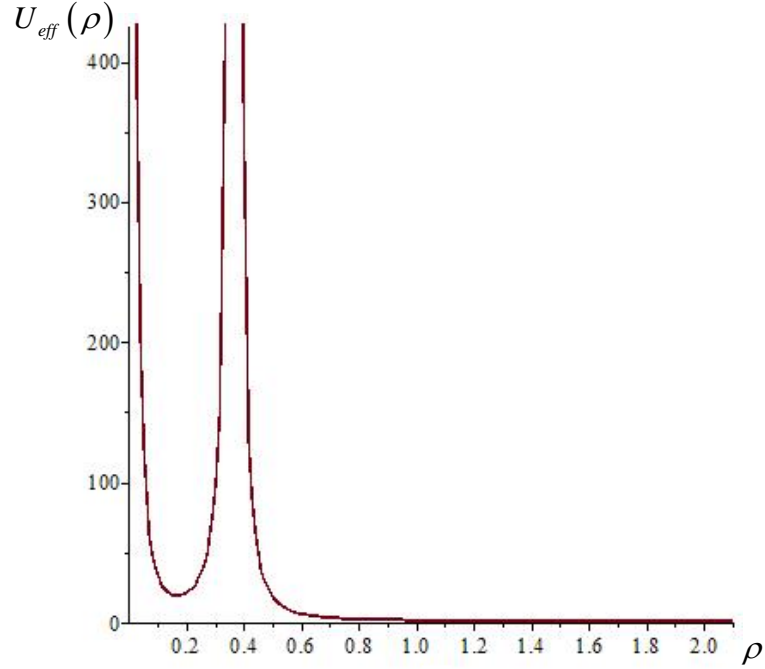

Fig.1. Behavior of the effective potential of the Dirac equation in the field of RN naked singularity as $\alpha = 0.25$, $\alpha_Q = 0.5$, $\alpha_{em} = 1$, $\kappa = -1$, $\varepsilon = 1.5$, $\rho_{cl} = 0.3675$.

One can see two regions $\rho < \rho_{cl}$ and $\rho > \rho_{cl}$ separated by a potential barrier impenetrable to particles. In the outer region, there are no stationary states of Dirac particles. In the inner region $0 < \rho < \rho_{cl}$, the existence of stationary bound states of half-spin particles is possible.

**3.3** For opposite signs of charges $Q, e$ at certain values of initial parameters the typical form of the potential $U_{eff}(\rho)$ is given in fig. 2. The form of the potential does not change for the case of an uncharged Dirac particle either (see fig.3). Such $U_{eff}(\rho)$ relations testify to a possibility of existence of stationary bound states of half-spin particles in both cases.

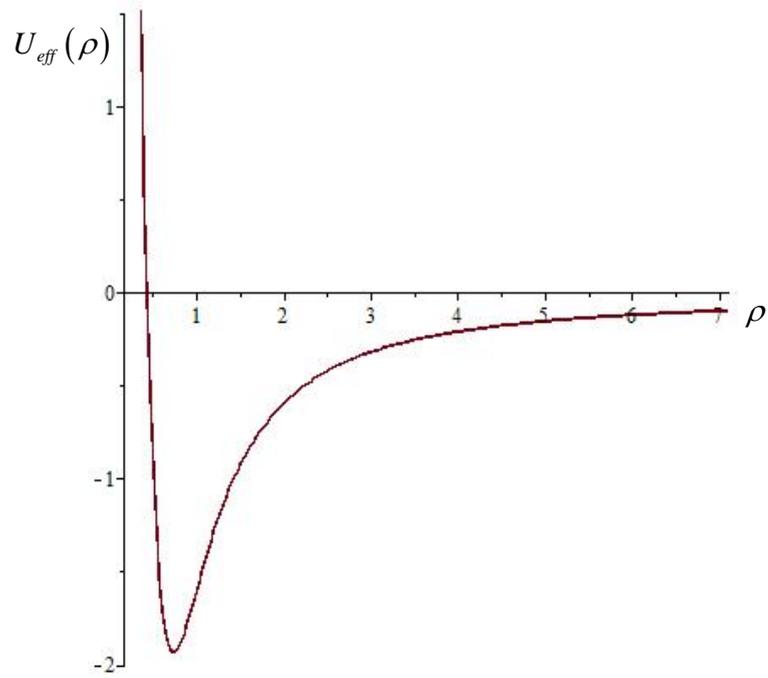

Fig. 2. Behavior of effective potentials of the Dirac equation in the field of RN naked singularity as $\alpha = 0.25,\ \alpha_Q = 0.5,\ \alpha_{em} = -1,\ \kappa = -1,\ \varepsilon = 0.6$

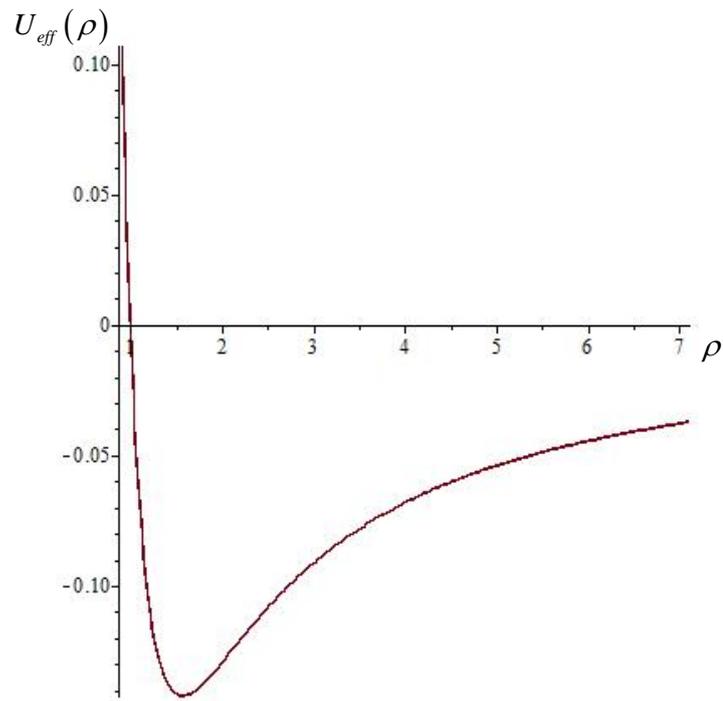

Fig. 3. Behavior of the effective potential of the Dirac equation in the field of RN naked singularity as $\alpha = 0.25,\ \alpha_Q = 0.5,\ \alpha_{em} = 0,\ \kappa = -1,\ \varepsilon = 0.98$.

## 4. Physical characteristics of interaction of half-spin particles with the field of Reissner-Nordström naked singularity

In the case of RN naked singularity $\alpha_Q^2 > \alpha^2$. Then, it follows from the relation (7) that

$$M < \sqrt{\alpha_{fs}} M_P \left|\frac{Q}{e}\right|, \quad (19)$$

$$m > \frac{M_P}{\sqrt{\alpha_{fs}}} \left|\frac{e}{Q}\right| \alpha. \quad (20)$$

For the charged Dirac particles of the Standard Model, the range of particle mass variations is from ~ 0,5 $MeV$ (an electron mass) to 170 $GeV$ (a $t$-quark mass). It is well-known that the constant of gravitational interaction in this case is extremely low as compared with the constant of electromagnetic interaction $\alpha_{em} = \alpha_{fs} \left|\frac{Q}{e}\right|$

$$\alpha < \left(3 \cdot 10^{-24} \div 10^{-18}\right) \left|\frac{Q}{e}\right|. \quad (21)$$

The energy spectra of bound states of the Dirac particles of the Standard Model in the field of RN naked singularity will, to a great extent, coincide with the electromagnetic spectra of appropriate hydrogen-like atoms.

In [10], the gravitational shift of the ground state of a hydrogen atom with the geometry of RN naked singularity was evaluated to be ~ $10^{-36}$ MHz.

Below, in order to demonstrate the existence of bound states of half-spin particles in numerical calculations three variants of atomic systems will be considered.

In these variants, the values $M, m, Q$ were selected for easy analysis of the computational results.

The first variant:

$$\alpha = 0.0018; \quad \frac{|\alpha_Q|}{\alpha} = 2; \quad \alpha_{em} = \frac{1}{137}. \quad (22)$$

The variant corresponds to the existence of the RN field source with $M = \frac{M_P}{2\sqrt{\alpha_{fs}}}$ and $Q = +e$; a particle with $m = M$ and with a charge $-e$ was chosen as a Dirac particle.

The second variant:

$$\alpha = 0.025; \quad \frac{|\alpha_Q|}{\alpha} = 2; \quad \alpha_{em} = 0,1. \tag{23}$$

The source of the RN field has $M = 0.585 M_P$ and $Q = +13.7e$; a particle with $m = 0.042 M_P$ and a charge $-e$ was chosen as a Dirac particle.

The third variant:

$$\alpha = 0.25; \quad \frac{|\alpha_Q|}{\alpha} = 2; \quad \alpha_{em} = 1. \tag{24}$$

The source of the RN field has $M = 5.85 M_P$ and $Q = +137e$.

A particle with $m = 0.042 M_P$ and a charge $-e$ was chosen as a Dirac particle. In all the variants, uncharged Dirac particles were also considered.

## 5. Determination of stationary bound states of half-spin particles in the field of Reissner-Nordström naked singularity

To solve the problem, we will use the system of equations (6) and Pruefer transformation [17]. Let us determine that

$$\mathrm{tg}\Phi(\rho) = \frac{F(\rho)}{G(\rho)}, \tag{25}$$

$$A^2(\rho) = F^2(\rho) + G^2(\rho). \tag{26}$$

Then,

$$\begin{aligned} F(\rho) &= A(\rho)\sin\Phi(\rho), \\ G(\rho) &= A(\rho)\cos\Phi(\rho). \end{aligned} \tag{27}$$

From the equations (6), one can derive an equation for the phase $\Phi(\rho)$ and an equation for $A(\rho)$

$$\frac{d\Phi}{d\rho} = \frac{1}{f_{R-N}}\left(\varepsilon - \frac{\alpha_{em}}{\rho}\right) + \frac{\cos(2\Phi)}{\sqrt{f_{R-N}}} - \frac{\kappa}{\rho\sqrt{f_{R-N}}}\sin(2\Phi), \tag{28}$$

$$\frac{d(\ln A)}{d\rho} = -\frac{1}{\rho f_{R-N}} + \frac{\alpha}{\rho^2 f_{R-N}} + \frac{\kappa}{\rho\sqrt{f_{R-N}}}\cos(2\Phi) + \frac{1}{\sqrt{f_{R-N}}}\sin(2\Phi). \tag{29}$$

## 5.1 Boundary conditions

To solve the equations (28), (29), it is necessary to specify boundary conditions as $\rho \to \infty$ and $\rho \to 0$ for the wave functions $F(\rho), G(\rho)$.

As $\rho \to \infty$, the asymptotic behavior of wave functions for finite motion is standard for centrally symmetric gravitational fields [18], [15].

As $\rho \to \infty$, the leading terms of the asymptotics are

$$F = Ce^{-\rho\sqrt{1-\varepsilon^2}},$$
$$G = -\sqrt{\frac{1-\varepsilon}{1+\varepsilon}} F, \qquad (30)$$

$$\text{tg}\Phi = -\sqrt{\frac{1+\varepsilon}{1-\varepsilon}}. \qquad (31)$$

As $\rho \to 0$, the asymptotic behavior of wave functions was identified in [10]. In our symbols

$$F(\rho) = f_0 \rho^{3/2} + f_1 \rho^{5/2} + f_2 \rho^{7/2} + \ldots$$
$$G(\rho) = g_0 \frac{1}{\rho^{1/2}} + g_1 \rho^{1/2} + g_2 \rho^{3/2} + \ldots \qquad (32)$$

Substituting (32) in the equation (6), one can derive recurrent relations for the coefficients $f_n, g_n$, including the relations between $f_0$ and $g_0$, lacking in [11]

$$\frac{f_0}{g_0} = \frac{|\alpha_Q| - \alpha_{em}}{2\alpha_Q^2}. \qquad (33)$$

It follows from (32), (33) that

$$\text{tg}\Phi = \frac{F(\rho)}{G(\rho)}\bigg|_{\rho \to 0} = \frac{|\alpha_Q| - \alpha_{em}}{2\alpha_Q^2} \rho^2 = 0, \qquad (34)$$

$$\Phi(\rho = 0) = k\pi, \quad k = 0,1,2\ldots \qquad (35)$$

## 5.2 A numerical method to solve the equations (28), (29)

In this paper, we use the following numerical method for solving the boundary problem (28), (31), (35). For the set of values $\varepsilon_i$, the Cauchy problem with a controlled initial condition as $\rho \to \infty$ (31) is solved numerically. The sought solution to the boundary problem corresponds to such a value $\varepsilon_i$, for which the condition at the origin is satisfied (35). In terms of the obtained results the applied method agrees with the alternative numerical method: first the Cauchy problem with the initial condition at the origin is solved (35) and then from $\varepsilon_i$ numbers those for which the condition at infinity (31) is satisfied are selected.

The equation (28) has a singularity at the origin, therefore it was being solved numerically till a controlled low value $\rho_0 = 10^{-4} \div 10^{-8}$.

To solve the Cauchy problem, the fifth-order Runge-Kutta implicit method with step control is used (the Ehle scheme of Radau IIA three-stage method [19]).

Based on the known function $\Phi(\rho)$, the function $A(\rho)$ is derived from the equation (29) by the direct numerical integration. Such a method leads to the correct result for solutions to a boundary- value problem (28), (31), (35).

### 5.3 Determination of the discrete energy spectrum

For three versions of $\alpha, \alpha_Q, \alpha_{em}$ values provided in item 4 with boundary conditions (30) - (35), discrete spectra of states of Dirac particles in the field of RN naked singularity were determined due to the solution to the equation (28). The computational results are represented in tables 1, 2. The values $(1-\varepsilon_n)$ were calculated with the accuracy to the first digit after the decimal point. The numerical values in the second digits after the decimal point can vary after special computations for convergence of mathematical results. In the given paper, the discrete spectra are provided for the qualitative evaluation without performing precision calculations.

Table 1. Numerical values $1-\varepsilon_n$ as the opposite signs of $Q, e$ charges for three variants of the values $\alpha, \alpha_Q, \alpha_{em}$

|  | $n=1$<br>$\kappa=-1$<br>$j=1/2$<br>$l=0$ | $n=2$<br>$\kappa=-1$<br>$j=1/2$<br>$l=0$ | $n=3$<br>$\kappa=-1$<br>$j=1/2$<br>$l=0$ | $n=2$<br>$\kappa=+1$<br>$j=1/2$<br>$l=1$ | $n=3$<br>$\kappa=+1$<br>$j=1/2$<br>$l=1$ |
|---|---|---|---|---|---|
| $\alpha=0.0018$<br>$\alpha_Q=0.0036$<br>$\alpha_{em}=-1/137$ | $4.14 \cdot 10^{-5}$ | $1.03 \cdot 10^{-5}$ | $4.53 \cdot 10^{-6}$ | $1.04 \cdot 10^{-5}$ | $4.58 \cdot 10^{-6}$ |
| $\alpha=0.025$<br>$\alpha_Q=0.05$<br>$\alpha_{em}=-0.1$ | $7.,85 \cdot 10^{-3}$ | $1.96 \cdot 10^{-3}$ | $8.61 \cdot 10^{-4}$ | $1.97 \cdot 10^{-3}$ | $8.71 \cdot 10^{-4}$ |
| $\alpha=0.25$<br>$\alpha_Q=0.5$<br>$\alpha_{em}=-1$ | $5.17 \cdot 10^{-1}$ | $1.96 \cdot 10^{-1}$ | $9.5 \cdot 10^{-2}$ | $3.18 \cdot 10^{-1}$ | $1.31 \cdot 10^{-1}$ |

|  | $n=2$<br>$\kappa=-2$<br>$j=3/2$<br>$l=1$ | $n=3$<br>$\kappa=-2$<br>$j=3/2$<br>$l=1$ | $n=3$<br>$\kappa=+2$<br>$j=3/2$<br>$l=2$ | $n=3$<br>$\kappa=-3$<br>$j=5/2$<br>$l=2$ |
|---|---|---|---|---|
| $\alpha=0.0018$<br>$\alpha_Q=0.0036$<br>$\alpha_{em}=-1/137$ | $1.03 \cdot 10^{-5}$ | $4.53 \cdot 10^{-5}$ | $4.58 \cdot 10^{-6}$ | $4.53 \cdot 10^{-6}$ |
| $\alpha=0.025$<br>$\alpha_Q=0.05$<br>$\alpha_{em}=-0.1$ | $1.94 \cdot 10^{-3}$ | $8.61 \cdot 10^{-4}$ | $8.7 \cdot 10^{-4}$ | $8.61 \cdot 10^{-4}$ |
| $\alpha=0.25$<br>$\alpha_Q=0.5$<br>$\alpha_{em}=-1$ | $2.19 \cdot 10^{-1}$ | $1.05 \cdot 10^{-1}$ | $1.09 \cdot 10^{-1}$ | $9.2 \cdot 10^{-2}$ |

Table 2. Numerical values $1-\varepsilon_n$ for an uncharged Dirac particle for three variants of the values $\alpha, \alpha_Q$

| | $n=1$<br>$\kappa=-1$<br>$j=1/2$<br>$l=0$ | $n=2$<br>$\kappa=-1$<br>$j=1/2$<br>$l=0$ | $n=3$<br>$\kappa=-1$<br>$j=1/2$<br>$l=0$ | $n=2$<br>$\kappa=+1$<br>$j=1/2$<br>$l=1$ | $n=3$<br>$\kappa=+1$<br>$j=1/2$<br>$l=1$ |
|---|---|---|---|---|---|
| $\alpha=0.0018$<br>$\alpha_Q=0.0036$<br>$\alpha_{em}=0$ | $1.62\cdot10^{-6}$ | $4.04\cdot10^{-7}$ | $1.7\cdot10^{-7}$ | $4.04\cdot10^{-7}$ | $1.75\cdot10^{-7}$ |
| $\alpha=0.025$<br>$\alpha_Q=0.05$<br>$\alpha_{em}=0$ | $3.1\cdot10^{-4}$ | $7.76\cdot10^{-5}$ | $3.45\cdot10^{-5}$ | $7.81\cdot10^{-5}$ | $3.5\cdot10^{-5}$ |
| $\alpha=0.25$<br>$\alpha_Q=0.5$<br>$\alpha_{em}=0$ | $2.3\cdot10^{-2}$ | $6.94\cdot10^{-3}$ | $3.23\cdot10^{-3}$ | $7.81\cdot10^{-3}$ | $3.5\cdot10^{-3}$ |

| | $n=2$<br>$\kappa=-2$<br>$j=3/2$<br>$l=1$ | $n=3$<br>$\kappa=-2$<br>$j=3/2$<br>$l=1$ | $n=3$<br>$\kappa=+2$<br>$j=3/2$<br>$l=2$ | $n=3$<br>$\kappa=-3$<br>$j=5/2$<br>$l=2$ |
|---|---|---|---|---|
| $\alpha=0.0018$<br>$\alpha_Q=0.0036$<br>$\alpha_{em}=0$ | $3.99\cdot10^{-7}$ | $1.,7\cdot10^{-7}$ | $1.75\cdot10^{-7}$ | $1.8\cdot10^{-7}$ |
| $\alpha=0.025$<br>$\alpha_Q=0.05$<br>$\alpha_{em}=0$ | $7.76\cdot10^{-5}$ | $3.45\cdot10^{-5}$ | $3.5\cdot10^{-5}$ | $3.45\cdot10^{-5}$ |
| $\alpha=0.25$<br>$\alpha_Q=0.5$<br>$\alpha_{em}=0$ | $7.47\cdot10^{-3}$ | $3.35\cdot10^{-3}$ | $3.5\cdot10^{-3}$ | $3.35\cdot10^{-3}$ |

Based on the computational results, let us note the existence of stationary bound states of uncharged Dirac particles arising due to the gravitational interaction alone. On the whole, at low values $\alpha, \alpha_Q, \alpha_{em}$ (the first and the second variants) the discrete spectra are similar to the spectra of hydrogen-like atoms with an ill-defined dependence of the binding energy $1-\varepsilon_n$ on the value of the orbital and total momentums of a particle. In the third variant, rather a strong coupling is ensured between the Dirac particles and the field of the RN naked singularity.

## 5.4 Radial wave functions

For illustration, figures 4 – 7 show the normalized probability densities $p(\rho) = \rho^2 \left( F^2(\rho) + G^2(\rho) \right)$ obtained by the numerical solutions to the equation (29) for eigenfunctions of states $1S_{1/2} \left( n=1, \kappa=-1, l=0, j=1/2 \right)$, $2P_{1/2} \left( n=2, \kappa=1, l=1, j=1/2 \right)$. The relations were obtained for the three variants of the atomic systems represented in item 4.

For all the variants, both charged and uncharged Dirac particles were considered.

Upon the whole, the behavior of probability densities is of the same nature as when considering atomic systems in the Minkowski space. One can note that for $1S_{1/2}$ - states the Bohr radius $r_B = \frac{\hbar^2}{me^2}$ (in dimensionless units $\rho_B = \frac{1}{\alpha_{em}}$) for all the variants is close to the appropriate maximums of the probability density.

Variant 1: $\rho_B = 137$; $\rho_{p\max} = 110$ - fig. 4;

Variant 2: $\rho_B = 10$; $\rho_{p\max} = 7.7$ - fig. 6;

Variant 3: $\rho_B = 1$; $\rho_{p\max} = 1.18$ - fig. 7.

In all the variants for uncharged Dirac particles (purely gravitational interaction) the maximums of the probability density shift towards larger values of $\rho$.

For $2P_{1/2}$ -states, the appropriate maximums of the probability densities as compared with $1S_{1/2}$ -states are also located in the region of larger values of $\rho$.

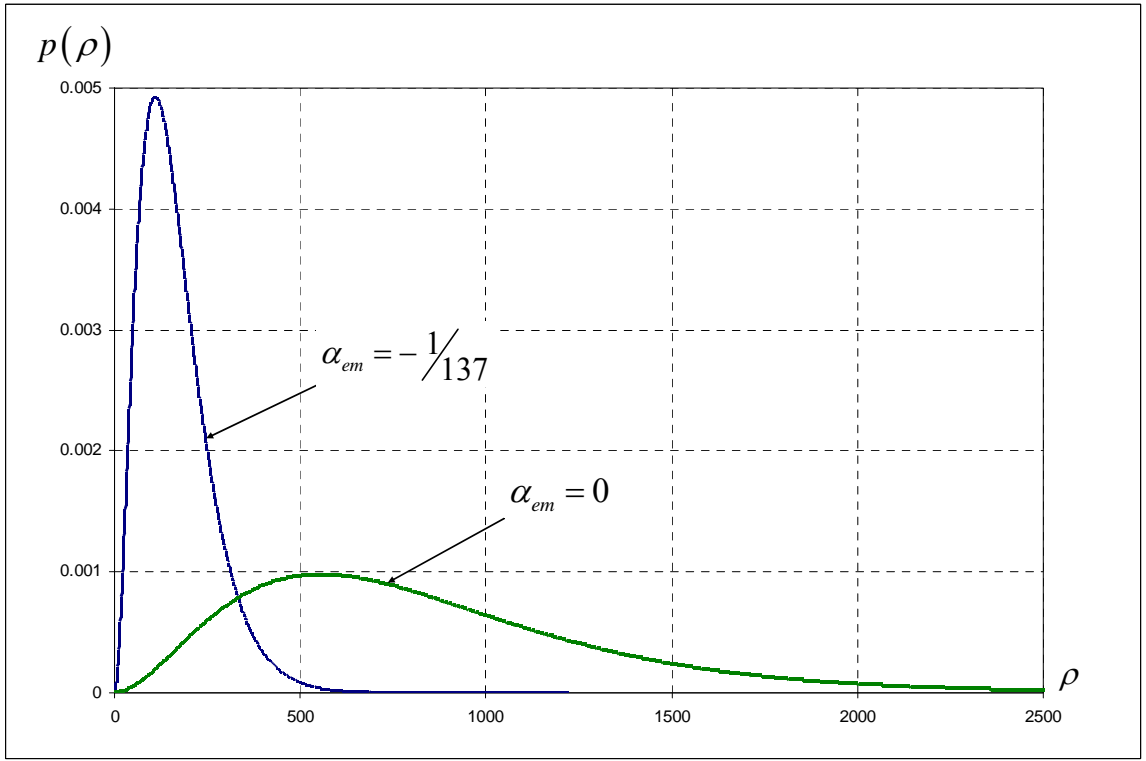

Fig. 4. The normalized probability density for $1S_{1/2}$-states of charged and uncharged Dirac particles as $\alpha = 0.0018$, $\alpha_Q = 0.0036$.

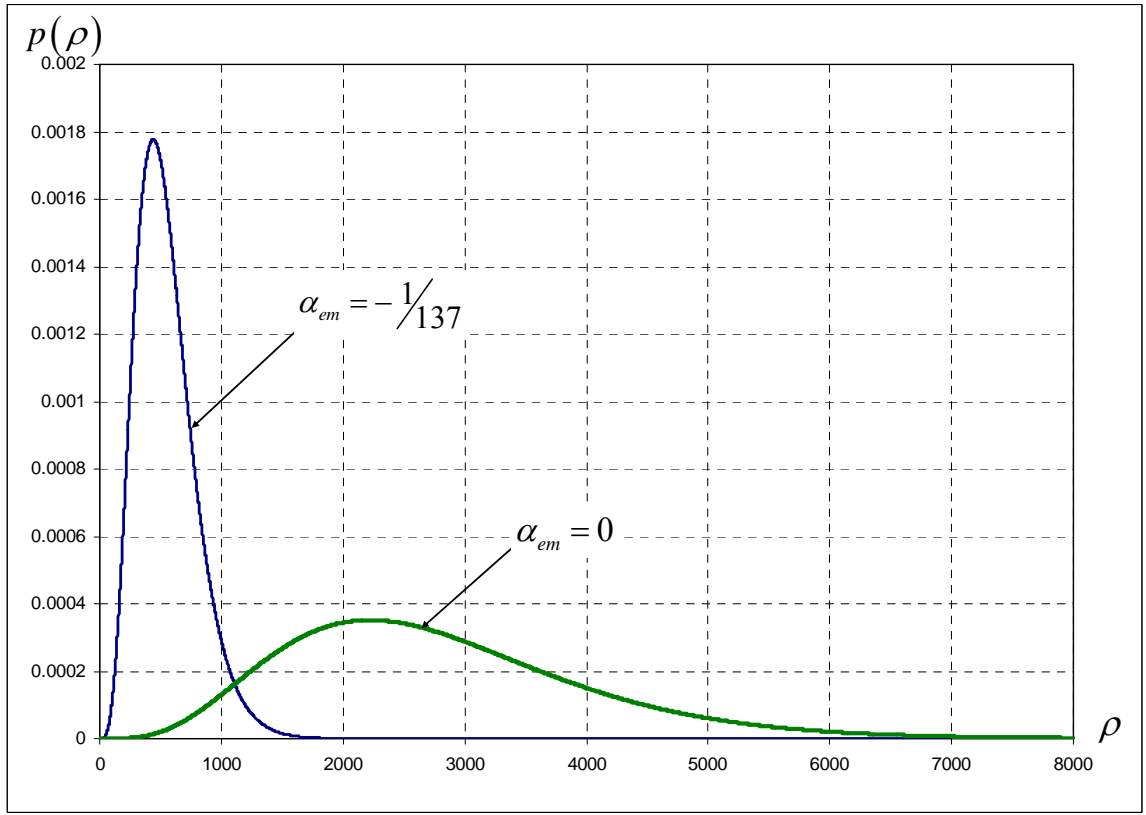

Fig. 5. The normalized density of probability for $2P_{1/2}$-states of charged and uncharged Dirac particles as $\alpha = 0.0018$, $\alpha_Q = 0.0036$.

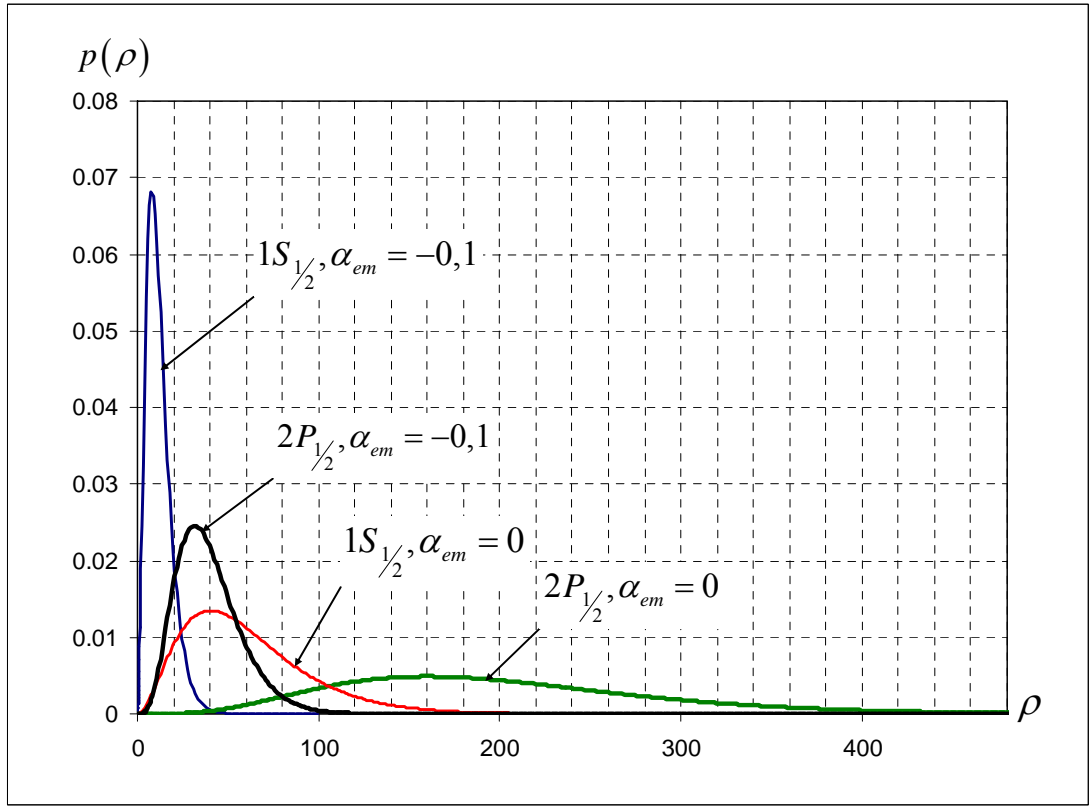

Fig. 6. The normalized probability density for $1S_{1/2}$, $2P_{1/2}$ - states of charged and uncharged Dirac particles as $\alpha = 0.025$, $\alpha_Q = 0.05$.

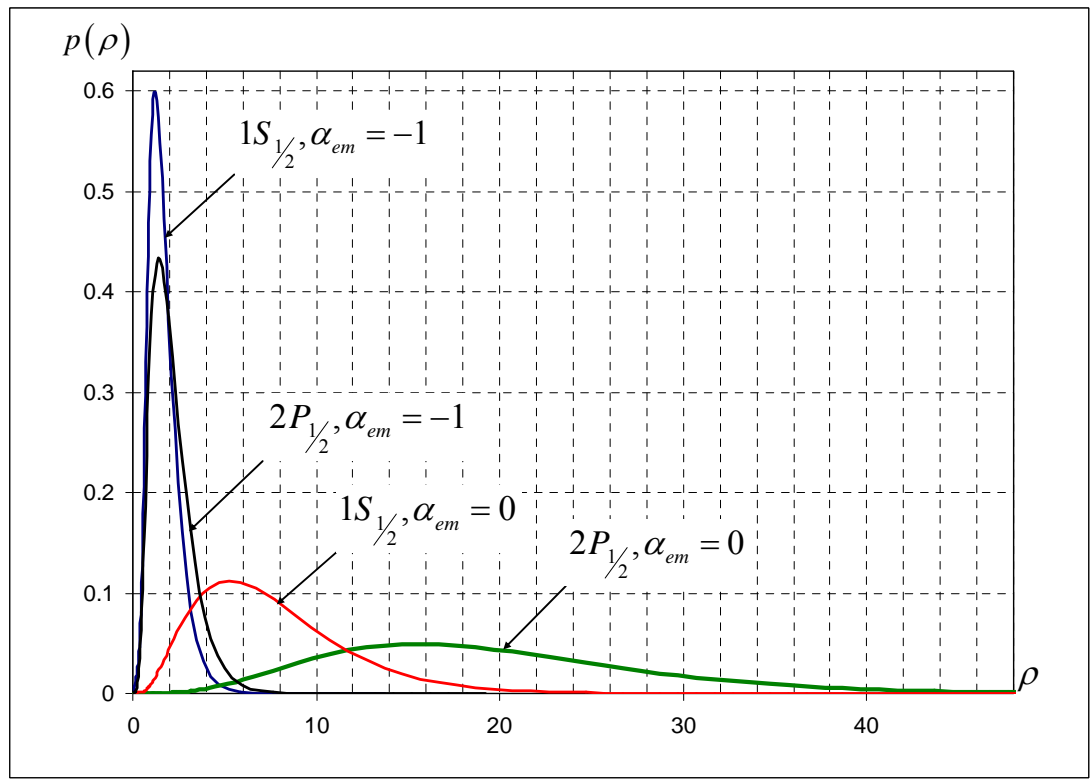

Fig. 7. The normalized probability density for $1S_{1/2}$, $2P_{1/2}$ - states of charged and uncharged Dirac particles as $\alpha = 0.25$, $\alpha_Q = 0.5$.

## 6. Discussions

The consideration of the quantum mechanics of the half-spin particle motion in the field of Reissner-Nordström (RN) naked singularity resulted in drawing the following conclusions:

- for any quantum –mechanical half-spin particle, irrespective of availability and sign of the electrical charge, the RN naked singularity is separated by the infinitely high positive potential barrier $\sim \frac{3}{8}r^{-2}$. This agrees with the conclusions [8] concerning the motion of spinless particles in the field of certain singular metrics.

As it is noted in [8], the availability of a repulsive barrier shielding the singularity not be seen as a threat to the cosmic censorship.

- with like charges of a Dirac particle and the source of RN naked singularity near the origin there exists the second completely impenetrable positive potential barrier $\sim \frac{A}{(r-r_{cl})^2}$. For elementary charges of a particle and the source of RN naked singularity as $\alpha_{em} \gg \alpha$, $\alpha_{em} \gg \alpha_Q$ and as particle energy $E \sim mc^2$, the value $r_{cl}$ is close to half a classical radius of a charged particle $r_e = \frac{e^2}{mc^2}$. As the value of the particle energy $E \gg mc^2$, the radius $r_{cl}$ decreases inversely to $E\left(r_{cl} = \frac{r_e}{E}\right)$.

- the analysis of the effective potentials and the direct numerical solutions to the Dirac equation have shown that the stationary bound states of half-spin particles in the field of RN naked singularity can exist in case of opposite charges of particles and the RN field source. In case of like charges, the bound states of a Dirac particle can exist under the potential barrier in the region $0 \leq r \leq r_{cl}$. In the region $r > r_{cl}$, there is no bound state of half-spin particles at the explored values of initial parameters.

- the stationary bound states of uncharged Dirac particles bound by the forces of gravitational interaction alone were obtained by direct computations of the Dirac equation in the field of RN naked singularity.

Let us pay attention to the difference in the dynamics of classical and quantum half-spin particles. It is well-known that in the classical case there appears the geodesic incompleteness of the RN naked singularity. The geodesics cannot pass across the surface whose radius is equal to half a classical radius of the source of RN naked singularity: $r_{cl} = \frac{Q^2}{2Mc^2}$. However, in quantum mechanics, depending on sings of charges, a Dirac particle can either be reflected from the repulsive barrier (14) or be in one of stationary bound states.

## **Acknowledgements**

The authors would like to thank A.L. Novoselova for the essential technical support while elaborating the paper.